# Vibrational Entropic Stabilization of Layered Chalcogenides: From Ordered Vacancy Compounds to 2D Layers

Roberto Prado-Rivera[1], Daniela Radu[1], Vincent H. Crespi[2,3], Yuanxi Wang[2,4]

[1] Department of Mechanical and Materials Engineering, Florida International University, Miami, FL 33199, USA

[2] 2-Dimensional Crystal Consortium, Materials Research Institute, Pennsylvania State University, University Park, Pennsylvania 16802

[3] Department of Physics, Department of Chemistry, and Department of Materials Science and Engineering, Pennsylvania State University, University Park, Pennsylvania 16802, USA

[4] Department of Physics, University of North Texas, Denton, Texas 76201, USA

**Abstract.**

Despite the rapid pace of computationally and experimentally discovering new two-dimensional layered materials, a general criteria for a given compound to prefer a layered structure over a non-layered one remains unclear. Articulating such criteria would allow one to identify materials at the verge of an inter-dimensional structural phase transition between a 2D layered phase and 3D bulk one, with potential applications in phase change memory devices. Here we identify a general stabilization effect driven by vibrational entropy that can favor 2D layered structures over 3D bulk structures at higher temperatures, which can manifest in ordered vacancy compounds where phase competition is tight. We demonstrate this vibrational-entropy stabilization effect for three prototypical ordered vacancy chalcogenides, $ZnIn_2S_4$ and $In_2S_3$, and $Cu_3VSe_4$, either by vacancy rearrangement or by cleaving through existing vacancies. The relative vibrational entropy advantage of the 2D layered phase originates mainly from softened out-of-plane dilation phonon modes.

## Introduction

The defining structural feature shared by all layered 2D materials is distinct atomic layers held together by weak non-bonded interactions. Yet *within* each layer, they exhibit a wide variety of structural motifs[1] such as honeycomb graphene, trigonal prismatic coordinated $MoS_2$, and square-netted FeOCl and CuBr. It is not easy to rationalize the stability of a layered structure in each case. For example, why does $MoS_2$ adopt a trigonal prismatic layered structure rather than other $AB_2$ lattices such as fluorite, pyrite, or rutile[2]? Three decades after the grand challenge of crystal structure prediction was raised by Maddox[3], modern global ground state search methods[4-6] can predict crystal structures and machine learning methods[7] can partially classify whether compounds prefer 3D or 2D bonding geometries, yet the underlying reason *why* a given structure is adopted often remains elusive. We propose vibrational entropy stabilization as a general mechanism to stabilize 2D layered structures. We further rationalize the stability of several classes of 2D layered materials by deriving them from ordered vacancy compounds (OVC) – 3D bulk materials that adopt simple structures (such as rocksalt or zincblende) with a ordered and stoichiometric fraction of sites left vacant. We do this by the two means: vacancy **rearrangement** or **cleaving**. In rearrangement, the point vacancies in an OVC are concentrated into vacancy planes, leaving separated layers. For example, a hypothetical ordered vacancy compound $A_3\square_1$ with one vacancy every four sites in a square lattice (Fig. 1) could arrange into a 2D trilayer material (left path). When cleaving, we split the structure along planes containing point vacancies (dashed circles in Fig. 1, right path) while removing atoms (crosses in Fig. 1) to leave separated layers.

We discuss three realistic prototypical OVCs where these two approaches apply, with $ZnIn_2S_4$ and $In_2S_3$ demonstrating vacancy rearrangement and $Cu_3VSe_4$ demonstrating cleaving. Each of these 3D OVCs is already on the verge of favoring a layered structure. $ZnIn_2S_4$ and $In_2S_3$ are both polymorphic, with bulk and layered polymorphs within ~0.1 eV/atom of each other and interconvertible by temperature or pressure[8, 9]. $Cu_3VSe_4$ occupies midpoint of the $Cu_4TiSe_4 \to Cu_3VSe_4 \to Cu_2MoSe_4$ series, where advancing metal valency yields a structural transition of 3D bulk → 3D bulk OVC → 2D layered. These mechanisms could be potentially generalized to carbides and nitrides with polymorphism[10, 11] such as the defective cubic system $TiC_{1-x}\square_x$ and the MXene $Ti_3C_2$.

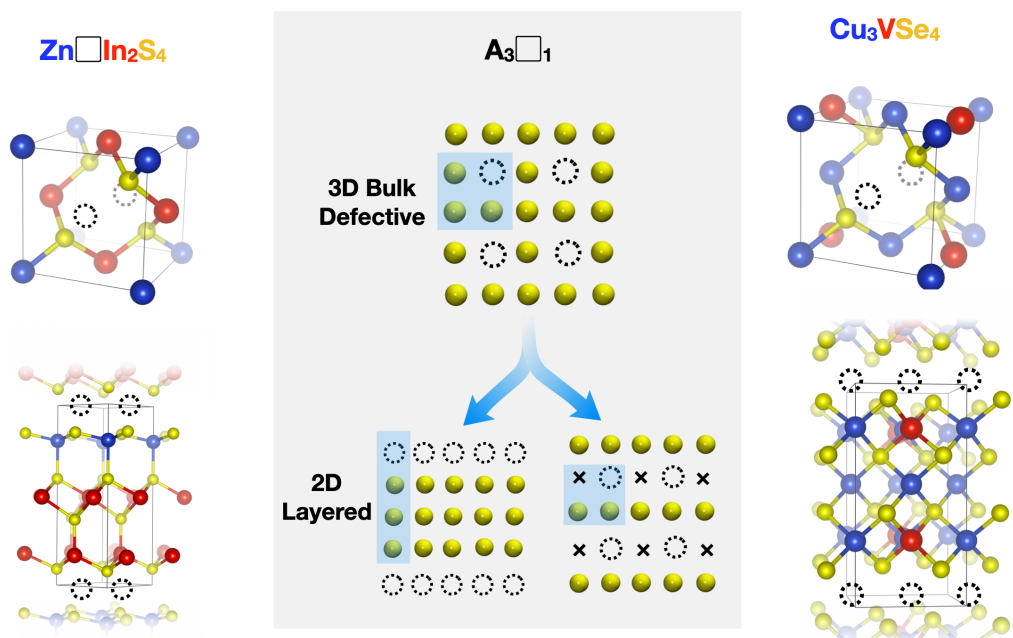

Figure 1. **Center panel:** Converting a bulk 3D ordered vacancy compound ($A_3\square_1$) into a 2D layered structure of the same stoichiometry by rearrangement or cleaving. **Side panels:** Structures of $ZnIn_2S_4$ and $Cu_3VSe_4$ in 3D bulk and 2D layered forms. Vacancies are marked by dashed circles.

Rational approaches to stabilizing layered chalcogenides may have far-reaching consequences in achieving passivated van der Waals surfaces for technologically important chalcogenide materials. Polymorphs of $ZnIn_2S_4$-type materials[9, 12, 13] are important in photodetectors, switches, photovoltaics and batteries[14]. The sulvanite-type compounds $Cu_3MX_4$ (M=V, Nb, Ta and X=S, Se) have promise in photovoltaic thin films[15-17] and photocatalysts for water splitting[18, 19]. Nanoscale morphologies have recently been realized for these compounds in the form of nanocrystals and nanosheets[20-25], with size-tunable intermediate bands (IB) that are present in both bulk and nanoscale structures[20-22]. The optical band gaps of $Cu_3V(S,Se)_4$ are especially favorable for solar-absorbing materials and have motivated attempts to achieve tunable band gaps by engineering nanoscale forms[20-26]. Furthermore, thin films made from Cu$_3$VSe$_4$ nanocrystals have demonstrated substantial stable *p*-type photocurrents[22].

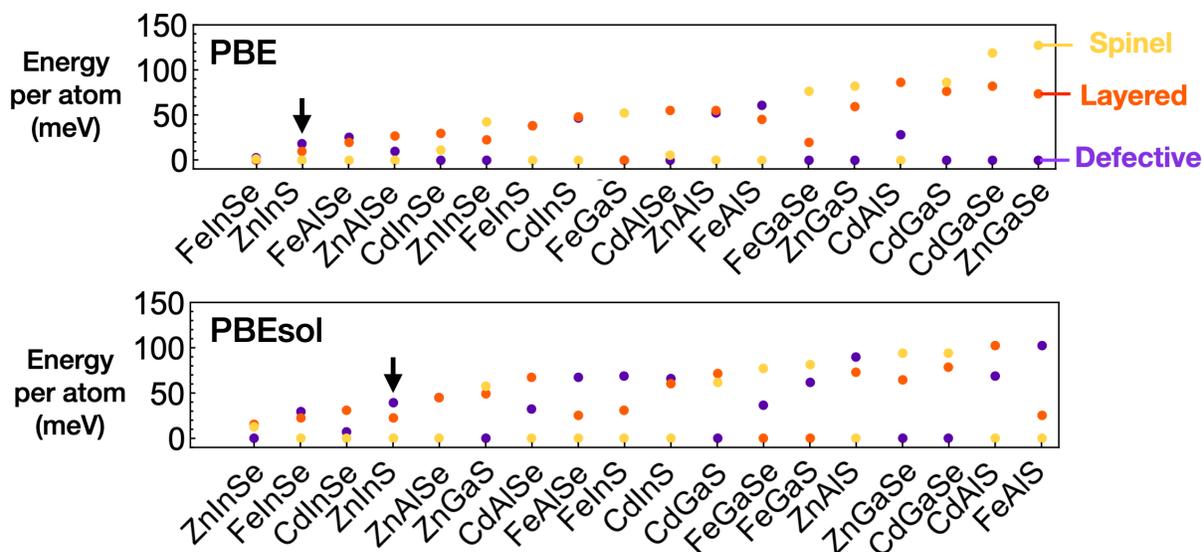

Figure 2. Per-atom DFT total energies for competing phases in 18 II–III$_2$–IV$_4$ materials systems, calculated with two exchange-correlation functionals in the upper and lower panels. ZnIn$_2$S$_4$ (black arrow) is chosen as our first focus.

## Results

We first consider the simpler of the two approaches, vacancy rearrangement, since stoichiometry is preserved. We choose a compound where the phase competition between bulk 3D and layered 2D is finely balanced, to within typical free energy scale of vibrational entropy contributions. A well-known structural family exhibiting phase competition between an ordered vacancy structure and a layered structure is the II–III$_2$–IV$_4$ family, which includes more than 30 compounds[9, 27]. Competing phases include defective zincblende (also called thiogallate), spinel, and layered phases. Fig. 2 shows the phase stabilities of 18 prototypical II–III$_2$–IV$_4$ compounds ranked by the closeness of the phase competition, from tightly competing at left to uncompetitive at right. All energetics are calculated within density functional theory using the PBE (upper panel) and PBEsol functionals (lower panel), both including semi-empirical van der Waals corrections under the DFT-D3 scheme. Both functionals perform reasonably well: PBE correctly predicts 13 of the 18 experimental ground-state phases while PBEsol predicts 11. For example, PBE predicts CdIn$_2$Se$_4$ to favor the defective phase over the spinel phase by ~10 meV/atom, while PBEsol finds the reverse order. Both functionals miss the stability of the layered phases of FeAl$_2$S$_4$, FeAl$_2$Se$_4$, and ZnIn$_2$S$_4$ and the defective phase of ZnAl$_2$Se$_4$. Overall, the PBE functional performs exceptionally well for the Cd and Fe compounds but struggles with the Zn compounds and appears to fare better than PBEsol (see Supplementary Notes 1 for a comprehensive comparison between predicted and experimentally realized phases).

ZnIn$_2$S$_4$ has one of the tightest phase competitions for both exchange-correlation functionals (arrows in Fig. 2); since it is also commonly reported in experiments[9, 27, 28], we chosoe to examine it in detail as a prototype. The ZnIn$_2$S$_4$ thiogallate structure (Fig. 1 top left) can be viewed as starting from ZnSe, a binary semiconductor following the octet rule, and substituting every four Zn around each Se with ZnIn$_2$ so that valency is preserved both on average and locally around each Se atom with one vacancy per formula unit, Zn☐In$_2$S$_4$. Alternatively, we can describe the ZnIn$_2$S$_4$ structure starting from pristine zincblende with four out of eight (4/8) tetrahedral holes occupied per formula unit and vacating one out of eight (1/8) of the tetrahedral holes. This description relates the thiogallate structure to the two other near-ground-state polymorphs of ZnIn$_2$S$_4$ – the spinel and layered[29] structures – in that they can all be considered to be under-occupied by 1/8 (see Supplementary Note 2). For the layered structure, the collection of these 1/8 vacant sites are of course just the interlayer galleries (Fig. 1 bottom left).

Experimentally, the thermodynamically stable ZnIn$_2$S$_4$ phase is the layered one, which seems to contradict the DFT results predicting spinel to be the ground state. We propose that this can be resolved by considering the larger vibrational entropy of the layered phase. The Mermin-Wagner Theorem formally predicts that long-range structural order is not possible in two dimensions due to an increased role for thermal fluctuations at lower dimensionality. Empirically, of course, the medium-range order that survives the theorem is generally sufficient to afford highly crystalline experimental samples of finite size[30]. We show below how vibrational fluctuations in two dimensions, which at small wavevector doom formal long-range order in the thermodynamic limit, actually *favor* the formation of layered structures in real materials due to their contribution to vibrational entropy at all wavevectors. To see this, we calculate the per-atom Gibbs free energy of the 3D bulk spinel (yellow) and 2D layered phases (red) of ZnIn$_2$S$_4$ in Fig. 3a, seeing the layered phase become more favorable in free energy above ~600 K. The bulk spinel free energy is used as a reference and is therefore set to zero at all temperatures. We attribute this crossover to reduced dimensionality softening phonons associated with layer dilation modes (or possibly shearing modes), which in turn increases vibrational entropy and thus reduces the free energy of the layered phase. Fig. 3b plots the cumulative phonon density of states (DOS) of for ZnIn$_2$S$_4$ in the bulk 3D spinel (yellow) and 2D layered (red) phase. The hump near 3 THz (red marker) indicates that phonons in that region have shifted to lower frequencies. We show the *cumulative* phonon DOS rather than the standard DOS (included in Supplementary Notes 3) because numerous van Hove singularities in the phonon DOS otherwise make it difficult to recognize the softened phonons. We can further inspect the cumulative vibrational free energy $\int_0^\omega F_{vib}\,\delta(\omega'-\omega_{i,k})\,d\omega'$ of the two phases in Fig. 3c, calculated for room temperature, where $F_{vib} = \sum_{i,k} k_B T \ln[1 - \exp(-\hbar\omega_{i,k}/k_B T)]$. Their difference is indicated by the shading between the two curves and also plotted as a separate gray curve at the bottom. One recognizes that the hump in Fig. 3b at 3 THz (vertical dashed line) is where the layered phase gains a head start towards a more favorable free energy. This lead remains near-constant as the integration proceeds to higher frequencies, since the vibrational free energy $\sum_{i,k} k_B T \ln[1 - \exp(-\hbar\omega_{i,k}/k_B T)]$ is dominated by lower-frequency contributions. The vertical dashed line at 3 THz actually marks the frequency of an out-of-plane intralayer dilation mode, where each layer dilates and contracts along *z* as shown by the phonon eigenvector in the inset of Fig. 3b. We will later show that the low-frequency hump always coincides with this intralayer dilation mode, since it originates from a van Hove singularity associated with this mode.

To demonstrate the generality of vibrational entropy stabilization, we consider another potential example, In$_2$S$_3$, before proceeding to a toy model that captures the essential lattice dynamics of layered phases. In$_2$S$_3$ is a member of the III$_2$–IV$_3$ family (III=Al/Ga/In, IV=S/Se/Te) where defective zincblende, defective spinel, and layered structures are common[31]. Its sibling In$_2$Se$_3$ is perhaps the most familiar to the 2D materials community for 2D ferroelectricity[32]. The tightly competing phases of In$_2$S$_3$ are more complicated. Here we focus on the two lowest-energy phases, the 3D bulk $\beta$ phase ground state and the layered $\alpha$ phase which is ~20 meV/atom higher for both PBE and PBEsol exchange-correlation functionals. The layered phase has been experimentally stabilized above 1023 K[31]. The two phases are structurally related by vacancy rearrangement. Similar to the ZnIn$_2$S$_4$ case, the cumulative phonon DOS for In$_2$S$_3$ in the bulk 3D $\beta$ (yellow) and layered (red) phases shown in Fig. 3b shows a hump near 4 THz (red marker), indicating phonons in that region soften in the layered phase. It follows that 4 THz (the vertical line in Fig. 3c) is where the layered phase establishes a lead in the cumulative vibrational free energy (shown for room temperature). This frequency also corresponds to an intralayer dilation mode in In$_2$S$_3$, as shown by the phonon eigenvector in the inset of Fig. 3b. The same vibrational-entropy stabilization effect can also occur in boron nitride[33], where low-frequency phonons in its layered hexagonal phase (h-BN) lead to a faster increase in vibrational entropy compared with the 3D cubic phase (c-BN), thus favoring the layered h-BN phase at temperatures higher than 335 K.

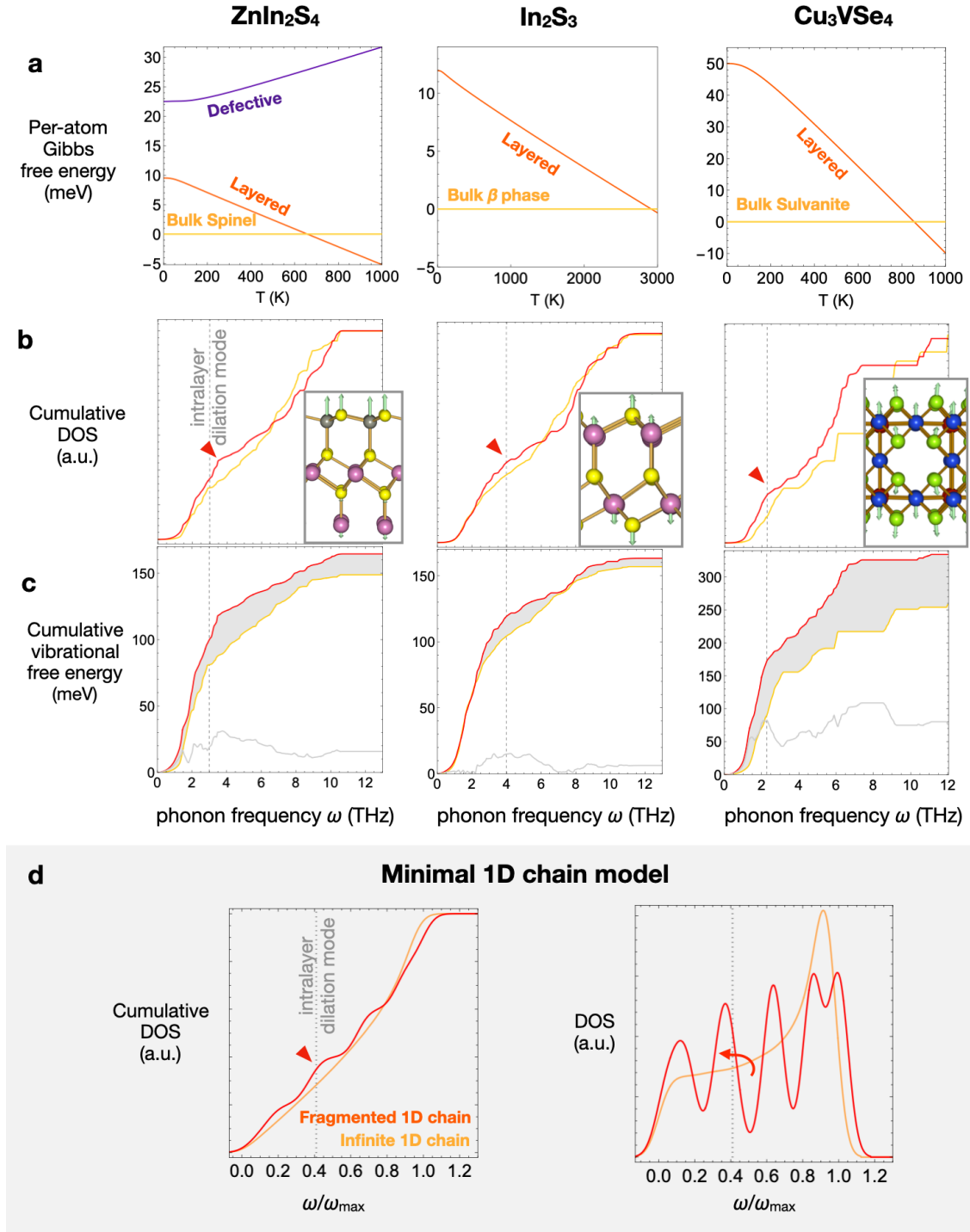

Figure 3. (a) Per-atom Gibbs free energy, (b) cumulative phonon DOS, (c) cumulative vibrational free energy, comparing bulk versus layered phases of (left column) ZnIn$_2$S$_4$, (center column) In$_2$S$_3$, and (right column) Cu$_3$VSe$_4$. The inset shows the intralayer dilation mode phonon eigenvector for three chalcogenide systems. In each case, the frequency of this mode is indicated by a vertical dashed line, and always coincides with a prominent hump feature in (c) marked by red triangles. In all cases, the favorability of the layered phase increases with increasing temperatures, due to phonons shifting to lower frequencies in layered phases. (d, left) Cumulative phonon DOS and (d, right) broadened phonon DOS for a minimal 1D 5-atom chain model that captures the difference between bulk and layered In$_2$S$_3$.

Now that we have established that the more negative vibrational free energies in layered ZnIn$_2$S$_4$ and In$_2$S$_3$ are due to softened phonons, next we develop a minimal model to show that the effect relates to layering generally and not material specificity. For a layered structure out-of-plane bonding along the $z$ direction is interrupted at every interlayer gallery, which can be modeled by a fractured 1D chain model along $z$ with, for example, five oscillators per segment to follow the In$_2$S$_3$ quintuple layer structure. The masses of the five oscillators follow the atomic masses of In and S in In$_2$S$_3$ (noting that the main results will not depend on the specific values of mass). A 3D bulk structure with strong bonding all along $z$ can be represented by an unbroken 1D chain. In Fig. 3d left panel, we calculate the same quantity as before – the cumulative phonon DOS – for this minimal 1D chain model. The yellow curve is for an uninterrupted 1D chain with constant nearest neighbor harmonic couplings. The red curve is for the same 1D chain but with every fifth coupling weakened by 80%, representing interlayer coupling, while all four other coupling strengths are boosted by 20%, representing the rest of the intralayer space being vacancy-free. The most prominent hump at low frequencies (second hump from the left, red marker) appears near the intralayer dilation mode (vertical dashed line), as can also be seen in the phonon DOS plot in the right panel of Fig. 3d, where we introduced a modest broadening to better visualize the softening phonons at low frequencies (red arrow). One also recognizes five distinct groups of eigenmodes, originating from the quintuple-layer structure. With increasing layer thickness (e.g. septuple layered ZnIn$_2$S$_4$), the spacing between these eigenmode groups would decrease, pushing the marked hump feature to smaller frequencies. This is consistent with the hump feature in ZnIn$_2$S$_4$ (~3 THz) being lower frequency than that in In$_2$S$_3$ (~4 THz). Although, this 1D model of course does not capture in-plane dispersion and material-specific phonon features, it does capture the essential phonon softening behavior that is seen in the cumulative phonon DOS for both ZnIn$_2$S$_4$ and In$_2$S$_3$. Thus, we conclude that phonon softening in layered phases can be mainly attributed to the introduction of weak interlayer couplings in layered phases, as compared to 3D bulk phases where the same vacancies are more evenly distributed.

What are the limits of vacancy rearrangement? Consider how the defective zincblende ZnIn$_2$S$_4$ structure may transform into the layered one. To maintain a 3:4 cation:anion ratio in the layered structure, the defective zincblende ZnIn$_2$S$_4$ needs to compact into a layered structure along the [111] direction with 3 cation (111) planes and 4 anion (111) planes, followed by a vacancy plane. The layered structure, now without vacancies, cannot condense into a thinner structure.

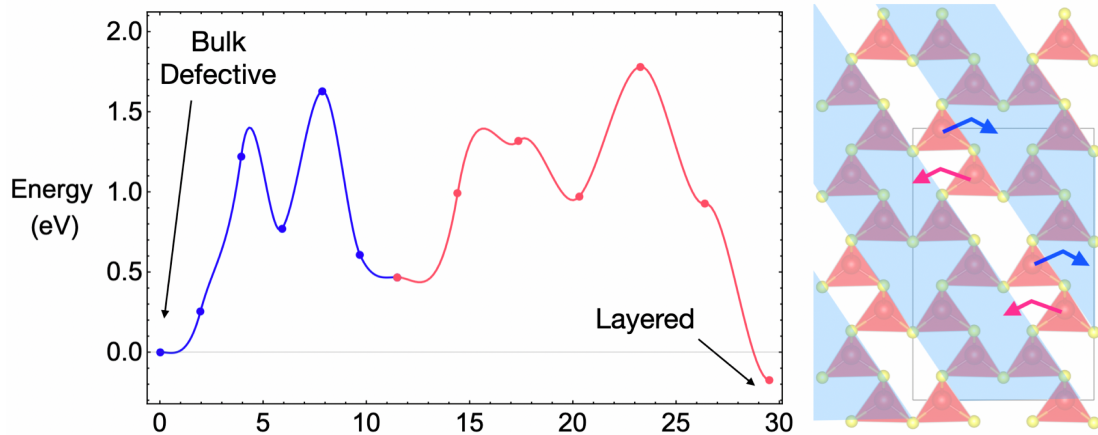

Figure 4. (Left) Energy profile of a possible phase transition reaction pathway between a bulk defective *Imm2* phase (right panel) and a layered phase (blue ribbons in right panel) of $In_2S_3$, obtained by nudged elastic band calculations. Blue and red segments in the energy profile correspond to same-colored atomic displacements in the right panel.

Similarly, for layered $In_2S_3$, the thinnest limit is 2 cation (111) planes and 3 anion (111) planes followed by a vacancy plane.

While we have established that competing phases of OVCs can be related by vacancy rearrangement, can these phases kinetically access each other? We briefly consider possible reaction pathways connecting OVC phases and layered phases. Although detailed atomic-scale kinetics is challenging to capture without molecular dynamics simulations over long length- and time-scales, we can estimate an upper bound for the kinetic barrier from a trial reaction pathway. We focus on the less structurally complex case of $In_2S_3$[34-37] and consider the phase transition between a metastable ordered vacancy zincblende *Imm2* phase[31] (labeled after its space group in Fig. 4, right panel) and the layered phase, with the goal of estimating a characteristic energy for the kinetic barrier to the transition between bulk defective and layered phases of chalcogenides. One such reaction pathway for $In_2S_3$ is shown in Fig. 4, where we move In atoms every three layers towards vacant sites in adjacent layers (arrows in Fig. 4). Two separate climbing image nudged elastic band[38] (CI-NEB) calculations were performed, each locating one saddle point for every atom displaced. The highest reaction barrier $E_b$ is ~1.5 eV, corresponding to a transition rate of $f \exp[-E_b/k_B T]$ = ~1 per second per to displace an In atom at 600 K, where we assume an attempt rate on the order of phonon frequencies, $f$ = 10 THz. This suggests a rapid defective→layered transition, provided that a thermodynamic driving force exists, e.g., at higher temperatures.

We now turn to the cleaving approach, using the $Cu_3VSe_4$ sulvanite structure as an example. The bulk sulvanite structure in Fig. 1 (top right) can be described as follows. Three Cu and four Se atoms occupy the usual zincblende sites (on two sublattices, each face-centered cubic). The V atom, surprisingly, resides at an edge-centered site[39] instead of the last face-centered site, occupying a tetrahedral hole where the tetrahedron shares an edge with its neighbors[40-42]. Compared to V placement at the vacant face-centered site, V atom placement at the edge-centered site has the same nearest-neighbor coordination environment (four Se), but a different next-nearest-neighbor environment (an additional six Cu). Pauling[39] argued that this V placement is more stable by appealing to his empirical electroneutrality principle for stability: the high-valency V requires more than four V-Se bonds (i.e. additional Cu-V bonds) to keep the charge on V closer to neutral (as defined within Pauling's framework of electronegativity, instead of modern charge partition methods). The vacancy that remains is not merely a nominal one – the channel opened by these vacancies allows high ionic mobility[43] of Cu in $Cu_3VS_4$. The metal atom layers along ⟨001⟩ alternate between a dense layer of $Cu_2V$ and a sparse layer of Cu.

Apart from effectively being an ordered vacancy compound, the $Cu_3VSe_4$ structure also involves Cu-Se antibonding, known in other Cu compounds to favor vacancy formation[44, 45]. This allows us to construct layered $Cu_3VSe_4$ structures by cleaving along the sparse Cu layers at low energetic cost. Such operations lead to surfaces terminated with a dense $Cu_2V$ layer followed by a Se surface layer (CuV-Se), labeled red in Fig. 5b. Before proceeding to discussing layered $Cu_3VSe_4$ structures, we will first verify that the CuV-Se termination is indeed the most stable one for constructing layers by systematically comparing this surface with all other energetically competing surfaces. The extra care taken here to construct $Cu_3VSe_4$ layers is because they have not been experimentally reported to form atomic layers, except for a recent report (Ref. 23) on the synthesis of $Cu_3VSe_4$ in the form of 10-nm-thick layers, which could consist of atomic layers. We consider three additional terminations where the final two layers are Se-CuV, Se-Cu, and Cu-Se. The structural relaxations of each slab are detailed in the Supplementary Note 4. The surface energy comparison of the four terminations depends on elemental chemical potentials. Under thermodynamic equilibrium, the chemical potentials of the three elements are constrained by Eq. (2) (see Methods), leaving two independent variables; here we choose $\mu_V$ and $\mu_{Cu}$. All three chemical potentials are bounded from above by their bulk total energies (at chemical potentials beyond these values elemental bulk is favored), forming a triangular region in $\mu_V$-$\mu_{Cu}$ space (Fig. 5a). The upper bounds of $\Delta\mu_V$ and $\Delta\mu_{Cu}$ are set to zero, while the lower bounds are –3.54 and –1.18 eV, respectively. The chemical potentials are also bounded by considering competing binary phases such as $VSe_2$ and $CuSe_2$, which further constrain the chemical potential range to the quadrilateral region ABCD. All surface energies are calculated using Eq. (1) (see Methods).

The calculated surface energies of the four $Cu_3VSe_4$ terminations along the path A-B-C-D-A in Fig. 5a are shown in Fig. 5b. The surface energies show no crossings within the entire chemical potential range, with decreasing surface energies in the sequence Se-CuV, Cu-Se, Se-Cu, to CuV-Se. This ordering of surface stabilities is related to their electronic structures, as discussed in Supplmentary Note 5. The CuV-Se surface being by far the most stable is consistent with our initial expectation. We further considered cases where stoichiometries are kept the same as bulk $Cu_3VSe_4$, with one extra Cu at the center vacancy or one extra Cu inserted at a vacancy site near the surface. These have surface energies of 0.04 and 0.03 eV/Å$^2$ respectively; as these lie between the CuV-Se and Se-Cu configurations, they are intermediate stable surfaces despite the additional Cu atoms.

Finally, constructing layered slabs with CuV-Se surfaces, we arrive at a septuple layer of $Cu_5V_2Se_8$ (Fig. 1 bottom right). This layered phase has a surface energy of 0.02 eV/Å$^2$, similar to a typical vdW interlayer adhesion energy[46, 47] of 0.02 eV/Å$^2$, and a metastability of +70 meV/atom, within the empirical threshold for synthesizability of 0.1 eV/atom. Following the previous examples of $ZnIn_2S_4$ and $In_2S_3$, Fig. 3a shows that the layered $Cu_5V_2Se_8$ phase can be entropically stabilized above ~850K. A familiar low-frequency hump can be seen near ~2 THz in the cumulative DOS of Fig. 3b, corresponding to the difference in the cumulative vibrational free energy between the layered $Cu_5V_2Se_8$ and bulk sulvanite $Cu_3VSe_4$ saturating near the same frequency. Just as for $ZnIn_2S_4$ and $In_2S_3$, 2 THz corresponds to intralayer dilation of $Cu_5V_2Se_8$ (inset of Fig. 3b).

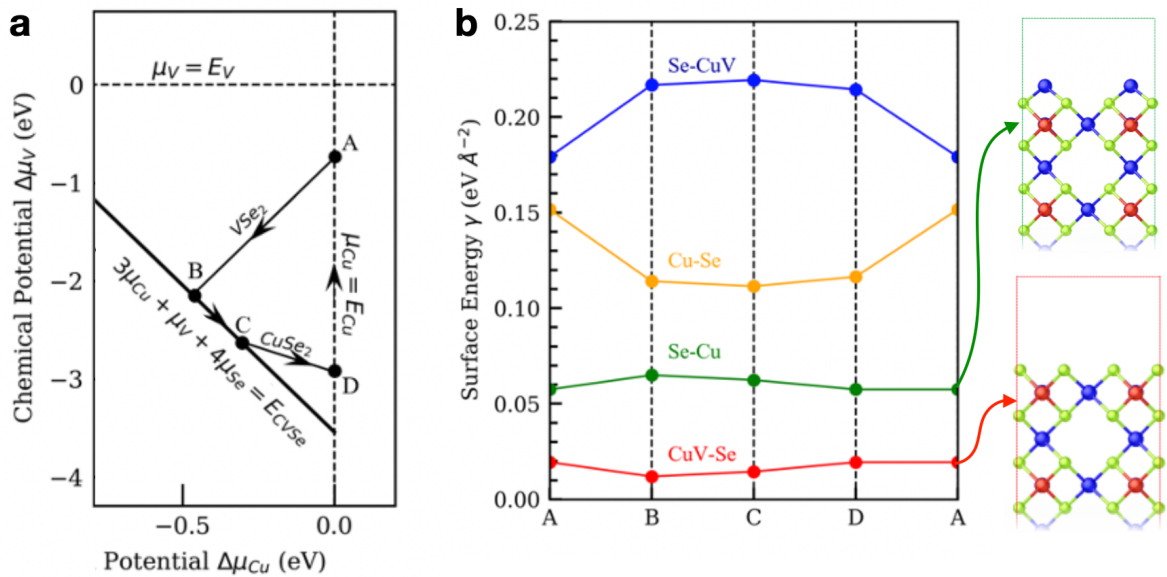

Figure 5. (a) Range of Cu and V chemical potentials $\Delta\mu_i = \mu_i - E_i^0$, where the reference energy $E_0$ is taken to be the ground-state bulk metal energy. The thick black line denotes the boundary of these chemical potentials under the assumption that the surface is in thermodynamic equilibrium with the underlying bulk material as given by the constraint $3\mu_{Cu} + \mu_V + 4\mu_{Se} = E_{bulk}$. The thinner black lines refer to the boundaries for the binary phases $VSe_2$ and $CuSe_2$. (b) Surface phase diagram of the $Cu_3VSe_4$ (001) surfaces plotted with respect to the chemical potential $\Delta\mu_{Cu}$, where the stable bulk phase energy has been taken as reference, along the pathway A-B-C-D-A (see arrows, left).

### Discussion

Further compacting $Cu_3VSe_4$ into a three-atom layer does not stabilize the structure: bringing every sparse Cu layer to its adjacent $Cu_2V$ plane results in a layered FeSe structure, which is +170 meV/atom above the ground state. This instability can be attributed to Cu tetrahedra having to share edges in the plane that is now $Cu_3V$, thus violating Pauling's third rule of structural stability (electrostatic repulsion between cations). This instability can be avoided in many ways, for example by replacing the $Cu_2V$ plane by by $Cu_2M_2$ (where M is a transition metal) or by $Cu_2M\square$. The former can be demonstrated by M=Mn, but one then realizes that $Cu_2Mn_2S_4$, or $CuMnS_2$, should be an anionic layer carrying a charge of −1 per formula unit and requiring counterions; indeed $RbCuMnS_2$ is a stable layered compound adopting the full FeSe structure. A more dilute intercalation of Rb can result in a $RbCu_2NbS_4$ phase. The same argument can be extended to describe the stability of layered $(K, Rb)_2TiS_4$ and presumably the $(K, Cs)_2(Mn, Cd)_3Te_4$ layered structure as well[48, 49]. The latter case ($Cu_2M\square$) is best demonstrated by $Cu_2WSe_4$, adopting a defective FeSe layered structure. However, since energetics cannot be fairly compared between $Cu_3VSe_4$ and $Cu_2WSe_4$, we consider the layered $Cu_5V_2Se_8$ structure obtained from the direct cleaving approach discussed above to be more useful for our discussions.

Potential generalization of the vibrational-entropy stabilization effect to other ordered vacancy compounds is possible, but needs careful consideration of disordered vacancy phases. In our examples the vacancy phases are all ordered, as constrained by local bonding environments and ensured by the presence of an electronic bandgap. For compounds with no bandgap, the competition between bulk defective phases and layered phases is less clear since the former can gain configurational entropy at higher temperatures. The is the presumably the case for MXenes such as $TiC_{1-x}$, where cubic disordered vacancy phases dominate at higher temperatures[50, 51].

In conclusion, we have identified a vibrational-entropy-driven stabilization effect that can favor 2D layered structures over 3D bulk ordered vacancy structures in compounds where phase competition is tight, with differences in per-atom energy on the order of thermal energies. We showed that this stabilization effect can transform OVC structures to layered ones either by rearrangements of vacancies into "vacancy planes" or by removing atoms in the same planes as the existing vacancies (i.e. cleaving at surfaces with low surface dangling bond densities). Vacancy rearrangement describes the transition from 3D bulk $ZnIn_2S_4$ and $In_2S_3$ to their layered phases. Cleaving describes the transition from 3D bulk sulvanite structure $Cu_3VSe_4$ to its layered form $Cu_5V_2Se_8$, with CuV-Se terminated (001) surfaces. In all three prototypical chalcogenide materials considered, layered phases can be stabilized at high temperatures by the relative vibrational entropy advantage of the layered phase, which originates mainly from soften phonons near the out-of-plane intralayer dilation phonon frequencies. This paradoxically inverts the standard reasoning of the Mermin-Wagner Theorem by deploying fluctuations in favor of layering, not to their detriment. The reaction kinetics of inter-dimensional structural transformation for the $In_2S_3$ case suggests that these such phase transitions can occur rapidly at relevant temperatures.

**Methods**

Total energy calculations and structural optimizations were performed using the PWscf (Plane-Wave Self-Consistent Field) code within the Quantum ESPRESSO package[52, 53]. The optimized lattice parameter $a$ = 5.566 Å of the bulk unit cell was obtained from total energy versus volume calculations fit to the Birch-Murnaghan equation of state[54], in agreement with previous studies[15, 42, 55]. We employ the Perdew-Burke-Ernzerhof exchange-correlation functional revised for solids[56, 57] (PBEsol) within the generalized gradient approximation[58] (GGA). Projector-augmented wave[59, 60] pseudopotentials for Cu ($3d^{10}4s^1$), V ($3d^34s^2$), and Se ($3d^{10}4s^24p^4$) were provided by the Pslibrary[61]. All the considered slab models were constructed by stacking three optimized unit cells along the [001] direction, aligned to the $z$-axis. A vacuum spacing of ~10 Å along the $z$-axis prevents spurious interactions between periodic images. Ionic relaxations were performed with a force threshold of $10^{-6}$ Ry/Bohr. The energy threshold for electronic self-consistency was $10^{-8}$ Ry. The Brillouin zone was sampled using a Monkhorst-Pack[62] grid of 5×5 for ionic relaxation and 8×8 for total energy calculations. The plane-wave kinetic energy cutoff for wavefunctions and charge densities was set to 60 Ry and 480 Ry, respectively. Gaussian smearing of 0.01 Ry was used on electron occupations.

Surface energies are calculated following the standard expression

$$\gamma = \frac{1}{2A}\left(G_{\text{surf}} - \sum_i N_i \mu_i\right) \tag{1}$$

where $G_{\text{surf}}$ is the free energy of the slab, and $N_i$ and $\mu_i$ are the number of atoms in the slab of species $i$ and their respective chemical potentials. The chemical potentials are constrained by

$$G_{\text{bulk}} = 3\mu_{\text{Cu}} + \mu_{\text{V}} + 4\mu_{\text{Se}} \quad (2)$$

where $G_{\text{bulk}}$ is the Gibbs free energy for bulk $Cu_3VSe_4$, since we assume the surface is in thermodynamic equilibrium with the underlying bulk crystal. The DFT total energies are incorporated by neglecting lattice vibration contributions so that the Gibbs free energy are approximated by DFT energies[63], $G \approx E^{\text{DFT}}$. The upper bounds for each chemical potential are set as the per-atom DFT total energy of the stable bulk phase[64]. The final chemical potential range is given in Fig. 5a.

## Acknowledgements

This work was supported by the National Science Foundation (NSF) through the 2-Dimensional Crystal Consortium, a Materials Innovation Platform under DMR-1539916 and DMR-2039351, the, and by the NSF FIU-2D Crystal Consortium Partnership for Research and Education in Materials (PREM) program under DMR-2122078. R. P.-R. acknowledges support from the NASA Graduate Fellowship Award. Y.W. acknowledges startup funds from University of North Texas and computational allocation on Texas Advanced Computing Center Lonestar6.

## Author contributions

Y.W. and R.P.-R. conceived and designed the theory calculations. R.P.-R. and Y.W. carried out the calculations and data processing. All authors wrote and reviewed the manuscript.

## Competing interests

The authors declare no competing interests.